\documentclass[pre]{revtex4}
\usepackage{amsmath}
\usepackage{color}
\usepackage{graphicx}
\newcommand{\tbR}{\tilde{\bf R}}
\newcommand{\bR}{{\bf R}}
\newcommand{\bP}{{\bf P}}
\newcommand{\bRa}{{\bf R}_{\rm a}}
\newcommand{\bRb}{{\bf R}_{\rm b}}
\newcommand{\br}{{\bf r}}
\newcommand{\bp}{{\bf p}}
\newcommand{\bF}{{\bf F}}
\newcommand{\tPhi}{\tilde{\Phi}}
\newcommand{\tbF}{{\tilde{\bf F}}}

\newcommand{\be}{\begin{equation}}
\newcommand{\ee}{\end{equation}}
\newcommand{\CVex}{C_V^{\rm ex}}
\newcommand{\CVexet}{C_{V,1}^{\rm ex}}
\newcommand{\CVexto}{C_{V,2}^{\rm ex}}
\newcommand{\Sex}{{S_{\rm ex}}}
\newcommand{\Sid}{S_{\rm id}}
\newcommand{\Fex}{F_{\rm ex}}
\newcommand{\Fid}{F_{\rm id}}
\newcommand{\pid}{p_{\rm id}}

\begin{document}
\title{Simplicity of condensed matter at its core: Generic definition of a Roskilde-simple system}
\author{Thomas B. Schr{\o}der}\email{tbs@ruc.dk}
\author{Jeppe C. Dyre}\email{dyre@ruc.dk}
\affiliation{DNRF Centre ``Glass and Time", IMFUFA, Department of Sciences, Roskilde University, Postbox 260, DK-4000 Roskilde, Denmark}
\date{\today}

\begin{abstract}
The theory of isomorphs is reformulated by defining Roskilde-simple systems (those with isomorphs) by the property that the order of the potential energies of configurations at one density is maintained when these are scaled uniformly to a different density. If the potential energy as a function of all particle coordinates is denoted by $U(\bR)$, this requirement translates into $U(\bRa)<U(\bRb)\Rightarrow U(\lambda\bRa)<U(\lambda\bRb)$. Isomorphs remain curves in the thermodynamic phase diagram along which structure, dynamics, and excess entropy are invariant, implying that the phase diagram is effectively one-dimensional with respect to many reduced-unit properties. In contrast to the original formulation of the isomorph theory, however, the density-scaling exponent is not exclusively a function of density and the isochoric heat capacity is not an exact isomorph invariant. A prediction is given for the latter quantity's variation along the isomorphs. Molecular dynamics simulations of the Lennard-Jones and Lennard-Jones Gaussian systems validate the new approach.
\end{abstract}

\maketitle

\section{Introduction}

In regard to structure and dynamics, liquids or solids dominated by van der Waals or weakly ionic and dipolar interactions, as well as metals, have more regular behavior than condensed matter dominated by directional bonds (hydrogen or covalent bonds) or strong Coulomb forces \cite{ric65,tem68,rowlinson,chandler,barrat,deb05,I,II,bag10,han13,pra14,dyr14,abr14}. This old insight has recently been formalized and confirmed by computer simulations of several models systems \cite{dyr14}. Thus it has been shown that systems with strong virial potential-energy correlations -- a characteristic of the former class of systems -- have ``isomorphic'' curves in the condensed-matter region of the thermodynamic phase diagram, curves along which structure and dynamics in properly reduced units are invariant to a good approximation. This means that for many quantities the phase diagram becomes effectively one-dimensional, a property that rules out anomalies \cite{pon11}. The systems in question were first referred to as ``strongly correlating'' \cite{I}, but this name was often confused with strongly correlated quantum systems and now the term ``Roskilde-simple systems'' or just ``Roskilde systems'' is used \cite{mal13,pra14,fle14,hen14,pie14,dyr14,buc14,fer14,sch14,abr14}. A recent review of the isomorph theory was given in Ref. \onlinecite{dyr14}.

An important experimental signature of Roskilde-simple systems is that they obey power-law density scaling over limited density variations, i.e., that the relaxation time is a function of $\rho^\gamma/T$ where $\rho$ is the density, $T$ the temperature, and $\gamma$ the so-called density-scaling exponent \cite{rol05,flo11}. These systems also obey isochronal superposition by which is meant the property that the average relaxation time determines the entire relaxation-time spectrum \cite{nga05,roe13}. A further application of the isomorph theory is the fact that for Roskilde-simple systems the melting line is an isomorph, which explains the invariances along it of several quantities \cite{ubb65,IV,dyr13,ped13}.

A system of $N$ particles in volume $V$ is considered with number density $\rho\equiv N/V$. The theory of isomorphs refers to quantities given in so-called reduced units \cite{IV}. The length and energy units are $\rho^{-1/3}$ and $k_BT$, respectively, the time unit depends on the dynamics (Newtonian or Brownian). In terms of the particle coordinates the configuration vector is defined by $\bR\equiv (\br_1,...,\br_N)$; its reduced-unit version is given by $\tbR\equiv\rho^{1/3}\bR$. 
The original isomorph theory \cite{IV} defines two thermodynamic state points with density and temperature $(\rho_1,T_1)$ and $(\rho_2,T_2)$, respectively, to be {\it isomorphic} if the following condition is obeyed: Whenever two physically important configurations of the state points, $\bR_1$ and $\bR_2$, have the same reduced coordinates, i.e., $\rho_1^{1/3}\bR_1=\rho_2^{1/3}\bR_2$, the following applies

\be\label{isomeq}
\exp(-{U(\bR_1)}/{k_B T_1})\cong C_{12}\exp(-{U(\bR_2)}/{k_B T_2})\,.
\ee
It is understood that the constant $C_{12}$ does not depend on the configurations. Thus whenever two configurations of isomorphic state points have the same reduced coordinates, their canonical probabilities are (almost) identical. This implies (almost) identical structure and dynamics in reduced units \cite{IV}.

It was recently shown that the existence of isomorphs for a given system is conveniently expressed in the ``hidden-scale-invariance'' identity that factorizes the potential-energy function $U(\bR)$ as follows \cite{dyr13a,dyr14}

\be\label{hsi}
U(\bR)\cong h(\rho)\tPhi(\tbR)+g(\rho)\,.
\ee 
Here the function $\tPhi(\tbR)$ is dimensionless and state-point independent. Equation (\ref{hsi}) expresses a global, approximate scale invariance in the sense that the function $\tPhi(\tbR)$, which determines structure and dynamics in reduced coordinates, is unchanged for a uniform scaling of all particle coordinates. This global approximate scale invariance is in the present paper modified into a more local form of scale invariance valid along each isomorph separately. 

In practice, the theory proposed below does not drastically change the predictions of the original isomorph theory \cite{IV} although there are some subtle, but significant differences. In Sec. \ref{II} we present an intuitive approach emphasizing the underlying physical idea. Section III proceeds axiomatically and derives the isomorph theory from a new definition of Roskilde-simple systems. Because of its axiomatic approach Sec. \ref{III} can be read independently of Sec. \ref{II}. Section \ref{trehalv} establishes the connection between the two approaches and finally Sec. \ref{fire} summarizes the paper.

\section{Towards a generalized isomorph theory: An intuitive approach}\label{II}

For numerical tests it is convenient to transform Eq. (\ref{hsi}) into an equation relating the potential-energy surfaces at two different densities, $\rho_1$ and $\rho_2$. In the following we let $\bR_1$ and $\bR_2$ denote configurations at densities $\rho_1$ and $\rho_2$, respectively, which have identical reduced coordinates, i.e., $\rho_1^{1/3}\bR_1=\rho_2^{1/3}\bR_2\equiv\tbR$. By elimination of $\tPhi(\tbR)$ Eq. (\ref{hsi}) implies 

\be\label{hsi2}
U(\bR_2)\cong h_1(\rho_2)U(\bR_1)+g_1(\rho_2)\,.
\ee 
Here the functions $h_1(\rho_2)$ and $g_1(\rho_2)$ depend also on $\rho_1$, which henceforth plays the role of reference density for which reason the $\rho_1$ dependence is only indicated in the subscript $1$. Equation (\ref{hsi2}) describes how the potential-energy surface at density $\rho_1$ scales when density is changed to $\rho_2$, namely to a good approximation simply by a linear, affine transformation. In particular, Eq. (\ref{hsi2}) implies Eq. (\ref{isomeq}) if the temperatures involved obey $T_2/T_1=h_1(\rho_2)=h(\rho_2)/h(\rho_1)$ \cite{IV,dyr14}, which is thus the condition for identifying isomorphic state points. This observation forms the basis of the so-called ``direct isomorph check''\cite{IV} in which configurations drawn from an equilibrium simulation at $\rho_1$ are used to test the scaling by evaluating the potential energy after uniformly scaling the configurations to density $\rho_2$. As an example, in Fig.~1 this is done for the single-component Lennard-Jones (LJ) system with $\rho_1=1.0$ and $\rho_2=2.0$ (LJ units). The black and red points refer to drawing  $\bR_1$'s from equilibrium simulations at $T_1=2.0$ and $T_1=4.0$, respectively (LJ units). The original isomorph theory implying Eq.~(\ref{hsi2}) predicts these two scatter plots to lie on a common straight line. This applies approximately, but not exactly. Thus doubling the sampling temperature from 2.0 to 4.0, changes the estimated value of $h_1(\rho_2)$ by roughly 5\%. As demonstrated below, such small deviations have significant consequences for the variation of the isochoric heat capacity $C_V$ along the isomorphs \cite{bai13} and for the temperature dependence of the density-scaling exponent $\gamma$.

\begin{figure}[htb!]
\centering
\includegraphics[scale=0.4]{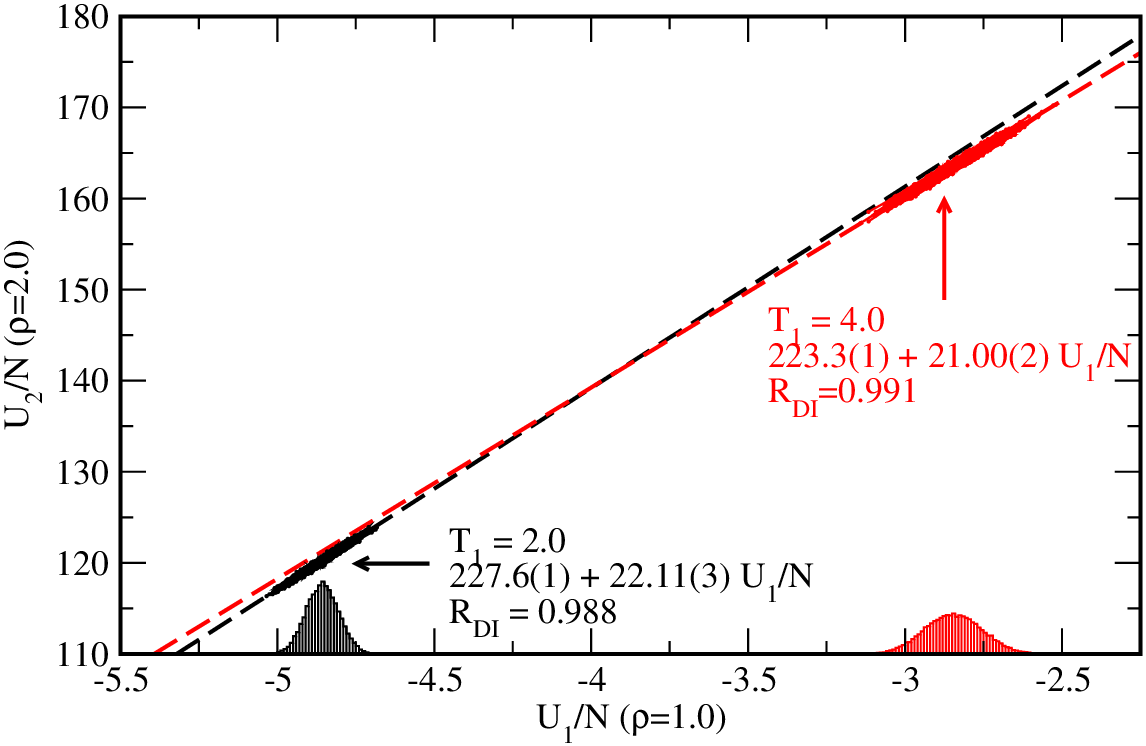}
\caption{Results from uniform scaling of configurations of a Lennard-Jones (LJ) liquid from density $1.0$ to density $2.0$ (in the LJ unit system defined by $\epsilon =\sigma =1$). Black gives a scatter plot for configurations generated at temperature $2.0$, red at temperature $4.0$, in both cases from simulations at the reference density $\rho_1$. Dashed lines are linear regression fits and numbers in parenthesis indicate the estimated error on the last digit. ${\rm R_{DI}}$ is the Pearson correlation coefficients for the two data sets. The strong correlations between original, $U_1\equiv U(\bR_1)$, and scaled potential energies, $U_2\equiv U(\bR_2)$, confirms that the LJ liquid is a Roskilde-simple system, i.e., one with strong virial potential-energy correlations and good isomorphs \cite{IV}. The distributions of $U_1$ for temperature 2.0 and 4.0, respectively, are indicated on the x-axis. 1728 LJ particles were simulated in the $NVT$ ensemble using a Nose-Hoover thermostat with time constant $0.2$. The time step was $0.001$ and the potential was cut and shifted at $4.5$. All simulations were carried out using the Roskilde University Molecular Dynamics (RUMD) code optimized for graphics processing units \cite{rumd}.
}
\label{Fig1}
\end{figure}

In order to generalize Eq. (\ref{hsi2}) to account for derivations from it, we assume a general one-to-one mapping of the potential-energy surface at $\rho_1$ to that at $\rho_2$:  
\be\label{feq}
  U(\bR_2)\cong f_1(\rho_2, U(\bR_1))\,.
\ee
The original formulation of the isomorph theory as expressed in Eq. (\ref{hsi2}) is recovered as the first-order Taylor approximation to Eq. (\ref{feq}). Consider a direct isomorph check corresponding, e.g., to the black points in Fig.~1 ($T_1=2.0$). For the relevant range of potential energies Eq.~(\ref{hsi2}) is evidently an excellent approximation to  Eq.~(\ref{feq}) if one identifies

\be\label{heq}
  h_1(\rho_2, U_1) \equiv 
   \left( \frac{\partial  f_1\left(\rho_2, U_1\right)}{\partial U_1} \right)_{\rho_2}
\ee
in which $U_1$ is the mean potential energy at the reference state point $(\rho_1,T_1)$. Defining $T_2 \equiv h_1(\rho_2, U_1) T_1$, the state point $(\rho_2,T_2)$ is \emph{isomorphic} to the state point $(\rho_1,T_1)$, compare the discussion above after Eq. (\ref{hsi}). Following Ref. \onlinecite{IV} it is straightforward to show that: 

\begin{enumerate}
\item The canonical probabilities of the configurations $\bR_1$ and $\bR_2$ are identical, implying that all structural characteristics are invariant in reduced units \cite{IV}; 
\item since the excess entropy depends only on structure, this quantity is also an isomorph invariant; 
\item the reduced forces associated with the configurations $\bR_1$ and $\bR_2$ are identical, which implies that the dynamics is isomorph invariant in reduced units. The predicted isomorph invariance of structure and dynamics for the state points of Fig. 1 is confirmed in Fig.~2.
\end{enumerate}

\begin{figure}[htb!]
\centering
\includegraphics[scale=0.4]{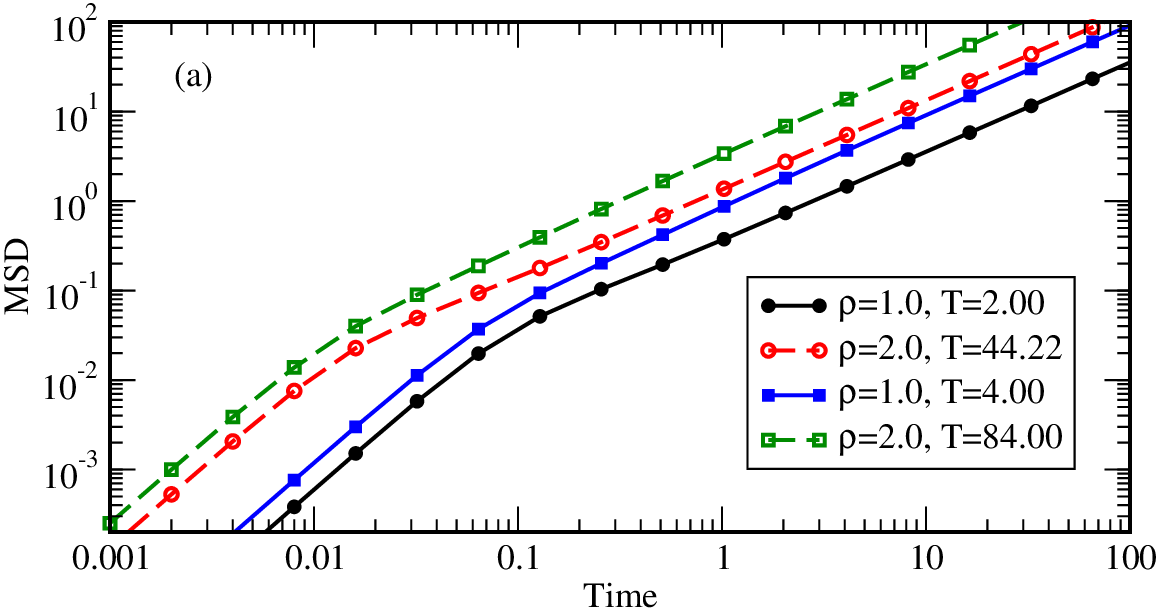}
\includegraphics[scale=0.4]{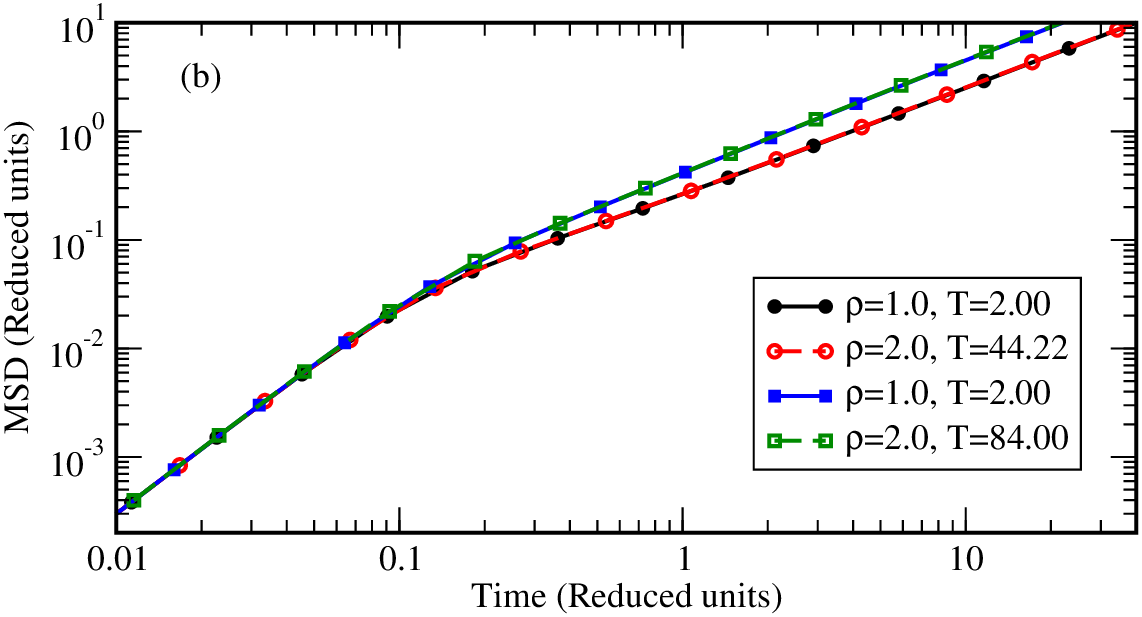}
\includegraphics[scale=0.4]{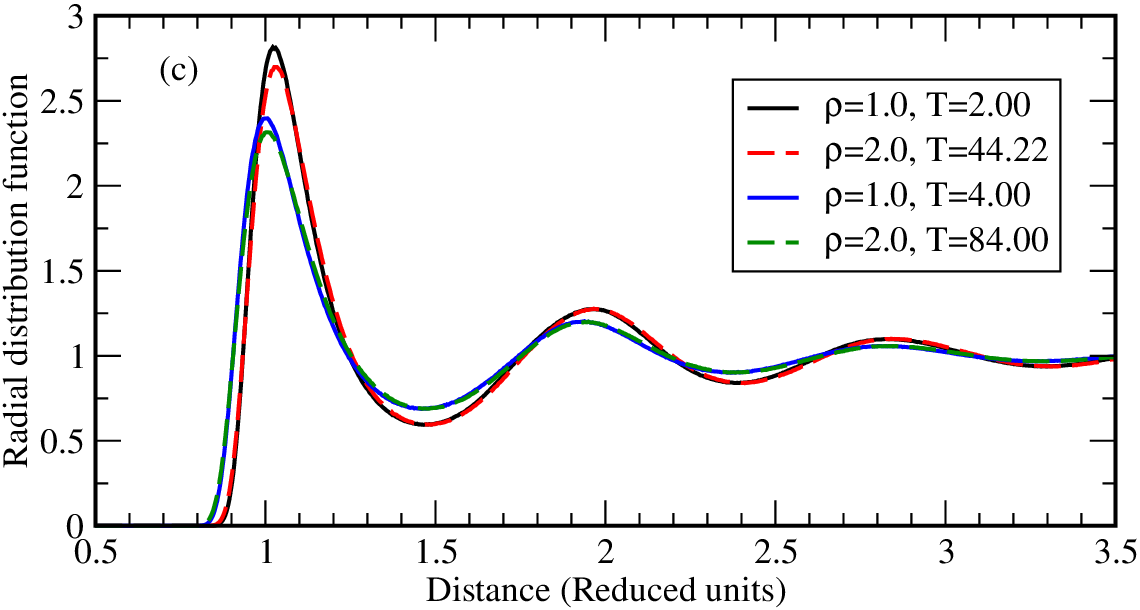}
\caption{Investigations of the isomorph invariance of structure and dynamics of the LJ system for the state points of Fig. 1 according to which $(\rho,T)=(1.0, 2.0)$ is predicted to be isomorphic with $(2.0, 2.0\cdot22.11)=(2.0,44.22)$ since the linear regression slope of the direct isomorph check (black points) in Fig. 1 is 22.11. Similarly, $(\rho,T)=(1.0, 4.0)$ is predicted to be  isomorphic with $(2.0,84.00)$. a) Mean-square displacement in standard LJ units for the four state points. b) Mean-square displacement in reduced units, demonstrating isomorph invariance \cite{IV}.  c) Radial distribution functions in reduced units, also demonstrating isomorph invariance though with minor deviations at the first peak maximum.}
\end{figure} 

An obvious question is: Are there corrections to these three points coming from the fact that they were derived from a first-order approximation to Eq.~(\ref{feq})? Based on considerations of the dependence on the system size $N$, this cannot be the case: The range of potential energies sampled at $(\rho_1,T_1)$ depends on the system size. The standard deviation of $U_1/N$ is proportional to $1/\sqrt{N}$, i.e., had we simulated a four times bigger system, the distributions in Fig.~1 would be half as wide. Thus approaching the thermodynamic limit, the first-order approximation to Eq.~(\ref{feq}) becomes better and better; in other words, the three above predictions are not influenced by the higher-order derivatives of Eq.~(\ref{feq}) since the predictions deal (implicitly) with the thermodynamic limit.

But which predictions \emph{do} change in the new formulation of the isomorph theory, Eq.~(\ref{feq})? In the formulation Eq. (\ref{hsi2}) the ratio $T_2/T_1$ is given by $h_1(\rho_2)$, which only depends on the densities involved, $\rho_1$ and $\rho_2$. In the new formulation this ratio is given by $h_1(\rho_2, U_1)$ and may also depend on the isomorph in question -- parameterized by $U_1$, the potential energy at the density $\rho_1$. It follows that the density-scaling exponent \cite{IV} $\gamma\equiv\left(\partial\ln T/\partial\ln\rho\right)_\Sex$ may vary on the isochores, whereas in the original isomorph theory $\gamma$ was predicted to be constant on these \cite{IV}. Fig.~3(a) shows that $\gamma$ indeed does change on the $\rho=1$ isochore, slowly approaching the limit $4$ valid at very high temperatures at which the LJ potential's repulsive $r^{-12}$ term completely dominates.

\begin{figure}[htb!]
\centering
\includegraphics[scale=0.4]{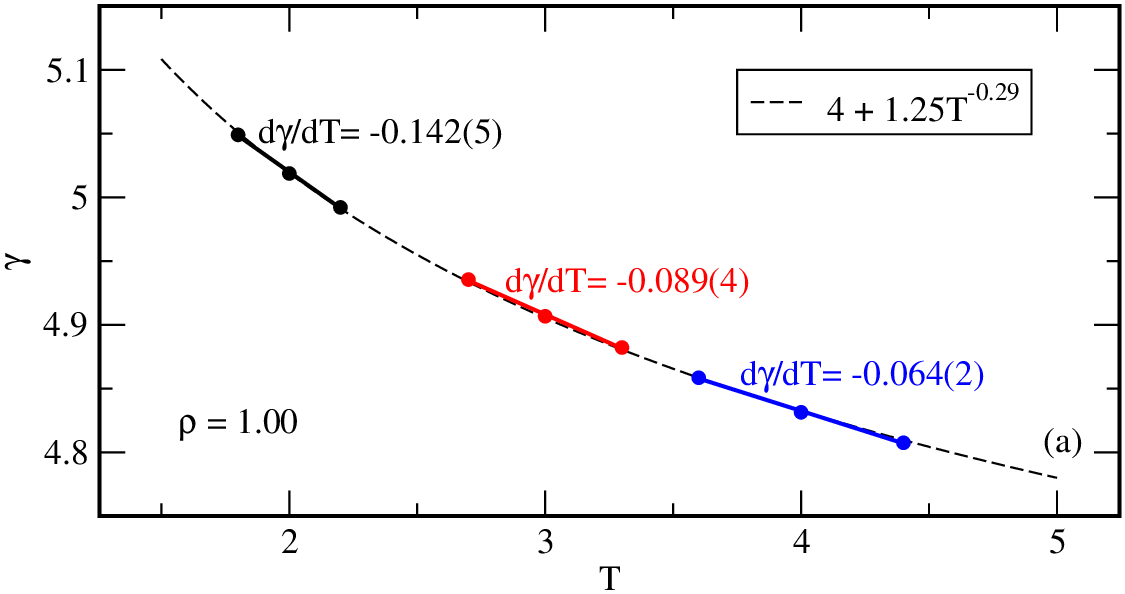}
\includegraphics[scale=0.4]{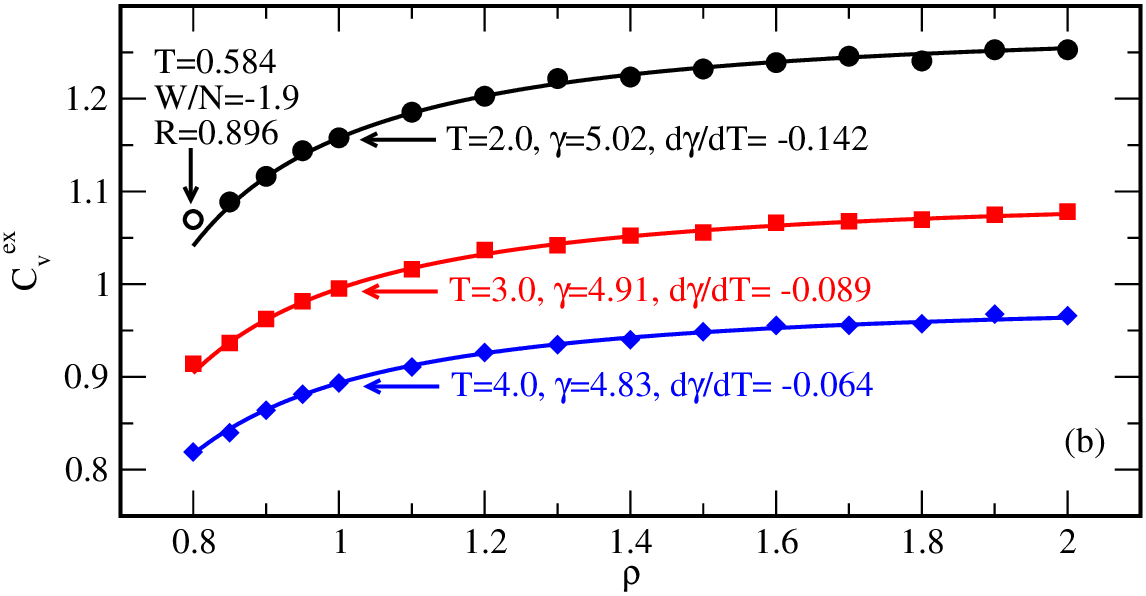}
\caption{
(a) Density-scaling exponent $\gamma$ of the LJ system calculated from the fluctuation expression \cite{IV} $\gamma=\langle\Delta W\Delta U\rangle/\langle(\Delta U)^2\rangle$ at density $1.0$ as a function of temperature, a quantity the original isomorph theory predicted to only depend on the density \cite{IV}. Dashed curve: a power law plus $4$ (the expected high-temperature limit) plotted as a guide to the eye. 
(b) Excess isochoric heat capacity along three different isomorphs, a quantity the original isomorph theory predicts to be isomorph invariant \cite{IV}. The full curves are the predictions of the new formulation of the isomorph theory (Eq. (\ref{cvex}) with $h_1(\rho_2) = (\rho_2/\rho_1)^4\left(\gamma_1/2 - 1\right) -  (\rho_2/\rho_1)^2\left(\gamma_1/2 - 2\right)$ in which $\gamma_1\equiv\gamma(\rho_1,T_1)$  \cite{boh12,ing12a}), based on input from the simulations at density $1.0$ in (a), i.e., without any fitting. As low densities are approached, the theory generally breaks down. For the lowest $T_1$ we included one data point where this starts to happen (open black circle). This point is a metastable liquid in the gas-liquid co-existence region, it has negative virial and a low virial potential-energy correlation coefficient $R$ -- these three properties all indicate breakdown of the isomorph theory \cite{IV}.}
\end{figure} 

Many thermodynamic response functions get a contribution from the {\it second} derivatives of $f_1(\rho_2,U_1)$  -- the exceptions being those for which the excess entropy is kept constant. An important case is the excess isochoric heat capacity, $\CVex$, which is predicted to be isomorph invariant in the original formulation of the theory \cite{IV} though this is not always accurately obeyed in simulations \cite{bai13}. Writing 
$\CVexto = \left({\partial U_2}/{\partial T_2} \right)_{\rho_2}=\left({\partial U_2}/{\partial U_1} \right)_{\rho_2}\left( {\partial U_1}/{\partial T_2} \right)_{\rho_2}$ and using $T_2=h_1(\rho_2, U_1) T_1$, it is straightforward to show that ${1}/\CVexto = {1}/\CVexet +  \left({T_1}/{h_1(\rho_2, U_1)}\right)\left({\partial{h_1(\rho_2, U_1)}}/{\partial U_1} \right)_{\rho_2}$. This can be rewritten as

\be
\CVexto = \CVexet \Big/\left[ 1 +  \left(\frac{\partial\,{\ln h_1(\rho_2, U_1)} }{\partial\ln T_1} \right)_{\rho_2}\right]\,. \label{cvex}
\ee
For LJ systems an analytical expression for $h_1(\rho_2, U_1)$ has been derived \cite{boh12,ing12a}, which combined with Eq.~(\ref{cvex}) shows that the variation of $\CVex(\rho_2)$ along an isomorph is determined by the two numbers $\gamma_1$ and  $\left(\partial\gamma_1/\partial T\right)_{\rho_1}$. Figure 3(b) tests the prediction for $\CVexto$ for three isomorphs generated with $T_1=2.0$, 3.0, and 4.0, respectively, at the reference density $\rho_1=1.0$. The values of  $\gamma_1$ and  $\left(\partial\gamma_1/\partial T\right)_{\rho_1}$ were determined from the $\rho=1.0$ simulations reported in Fig.~3(a). As can be seen in Fig. 3(b) the $\CVex(\rho_2)$ prediction agrees very well with the simulations.

\begin{figure}[htb!]
\centering
\includegraphics[scale=0.4]{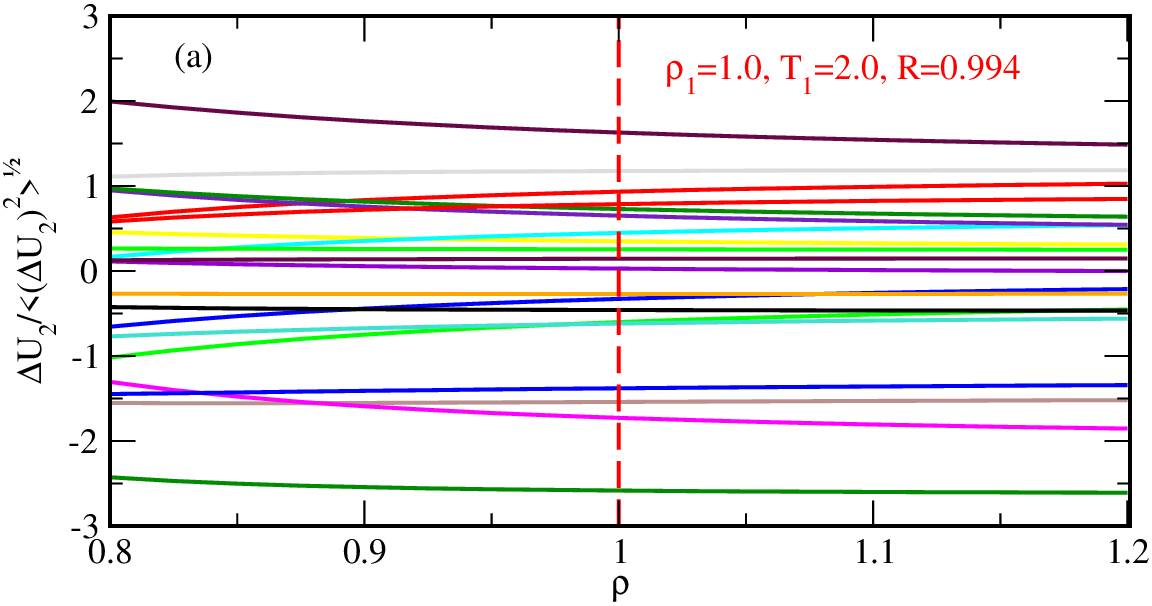}
\includegraphics[scale=0.4]{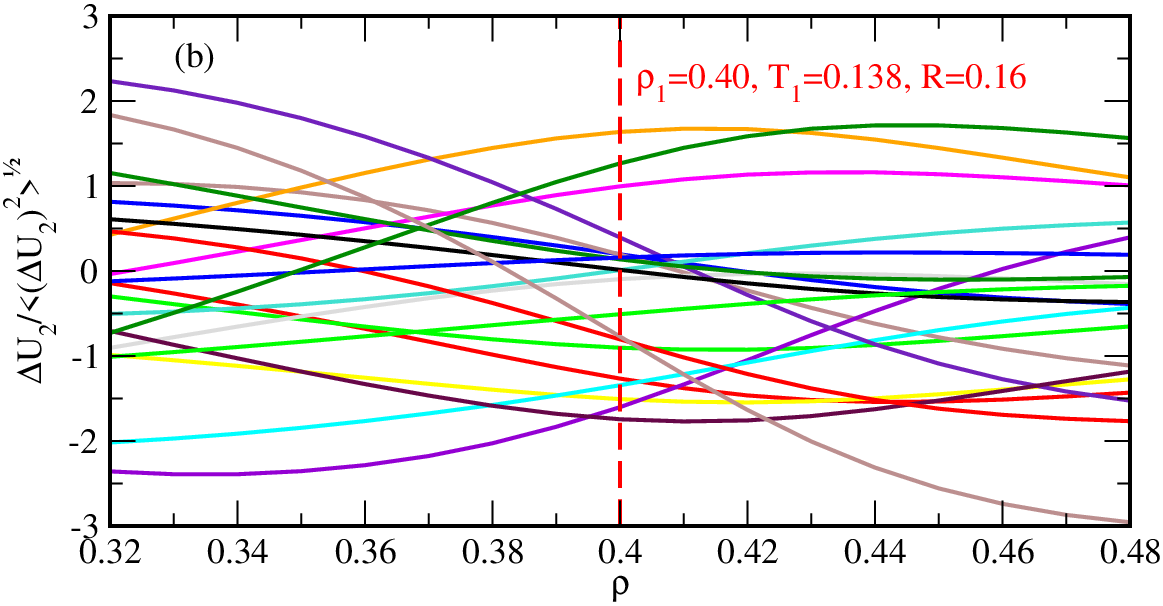}
\caption{Tests of the new definition of a Roskilde-simple system. Each figure shows the potential energies of 20 configurations taken from an equilibrium simulation, which have subsequently been scaled 20\% uniformly up and down in density and plotted as a function of density after being normalized by subtracting the average potential energy and scaled by the standard deviation (averages and standard deviations were determined from 1000 configurations). For a ideal Roskilde-simple system the curves cannot cross each other. 
(a) Data for the LJ system at the state point $(\rho,T)=(1.0, 2.0)$ where this system has strong virial potential-energy correlations ($R=0.99$).
(b) Data for the Lennard-Jones Gaussian (LJG) system \cite{deo06} at the state point $(\rho,T)=(0.4, 0.138)$ where $R=0.16$.}
\label{Fig_DP}
\end{figure}

\section{Axiomatic formulation}\label{III}

In the previous section an isomorph was identified by the potential energy at the reference density $\rho_1$, which is convenient in numerical tests of the theory. In this section we formalize the new theory, and this is based on identifying isomorphs by their excess entropy.

A Roskilde-simple system is henceforth defined by the property that whenever two configurations $\bR_a$ and $\bR_b$ refer to the same density, one has

\be\label{def}
U(\bRa)<U(\bRb)\,\,\Rightarrow\,\, U(\lambda\bRa)<U(\lambda\bRb)\,.
\ee
Because this will apply for scaling ``both ways'', an equivalent formulation is to replace $\Rightarrow$ by $\Leftrightarrow$, which in turn implies that if two configurations have the same potential energy, their scaled versions also have same potential energy. Thus an equivalent definition of a Roskilde-simple systems is 

\be\label{altdef}
U(\bRa)=U(\bRb)\,\,\Rightarrow\,\, U(\lambda\bRa)=U(\lambda\bRb)\,.
\ee
Any Euler-homogeneous potential-energy function obeys this condition, but so does a homogeneous function plus a constant; in view of this Eq. (\ref{altdef}) may be said to expresses a generalized homogeneity condition. Presumably no other systems obey Eq. (\ref{altdef}) for all configurations. In the following we make the weaker assumption that Eqs. (\ref{def}) or (\ref{altdef}) apply for {\it most} of the {\it physically relevant} configurations. This reflects the fact that the isomorph theory is inherently approximate for realistic models.

Figure \ref{Fig_DP} shows the potential energies as functions of density for scaled configurations of the LJ system, which has strong virial potential-energy correlations in the dense fluid phase, as well as for the Lennard-Jones Gaussian (LJG) system for which this is not the case. In each subfigure 20 configurations were picked from an equilibrium simulation at the density marked by the red vertical dashed line, and each of these were scaled uniformly to densities involving changes of $\pm 20$\%. According to Eq. (\ref{def}) curves giving the potential energy of such uniformly scaled configurations cannot cross each other. Since compression increases the potential energy dramatically, in order to facilitate comparison with Eq. (\ref{def}) we subtracted at each density the mean potential energy and scaled by the standard deviation -- still, a system is perfectly Roskilde-simple if no curves cross each other.  This is obeyed to a good approximation for the LJ system, but not for the LJG system; the low-density weak violations observed for the LJ system reflect the fact that it here gradually enters a region of weaker virial potential-energy correlations (compare Fig. 3(b)).

Below, in Secs. \ref{A} -- \ref{D} whenever a thermodynamic quantity is given without reference to a specific configuration $\bR$, this refers to the thermal equilibrium value at the thermodynamic state point in question. For instance, when we write $U$, this means the average potential energy at the state point in question, whereas writing $U(\bR)$ signifies the potential energy of the configuration $\bR$ (a different notation is used in Sec. \ref{trehalv} which relates the formal theory to Sec. \ref{II}). 

Recall that the entropy $S$ may be written as a sum of the ideal-gas entropy $S_{\rm id}$ at the same density and temperature and the so-called excess entropy, $S=S_{\rm id}+\Sex$. For an ideal gas $\Sex=0$; for any system with interactions $\Sex<0$ because no matter what is the nature of the interactions, such a system must be more ordered than an ideal gas. The Appendix reviews the definition of the excess free energy and other excess thermodynamic quantities; it also reviews the derivation of the microcanonical ensemble expression for $\Sex$, which is needed below for developing the new formulation of the isomorph theory.

In their theory of thermodynamic fluctuations Landau and Lifshitz define the entropy fluctuation at a given state point as the change in the equilibrium entropy ``formally regarded as a function of the exact value of the fluctuating energy'' \cite{lan58}. We shall adopt this definition of a microscopic entropy function, except for exclusively focusing on the {\it configurational} degrees of freedom, i.e., replacing energy by potential energy and entropy by excess entropy. More generally, given a system and a set of coarse-grained variables $A_1, ..., A_n$, for any given microstate $\Gamma$ one can define an entropy function $S(\Gamma)$ as the logarithm of the total number of microstates that have the same coarse-grained variables as $\Gamma$ \cite{rov14}. Clearly, $S(\Gamma)$ depends on the choice of coarse-grained variables. The case discussed by Landau and Lifshitz corresponds to that of a single coarse-grained variable, namely the energy; we here follow this except for using the potential energy.

We thus define the {\it microscopic excess entropy function} $\Sex(\bR)$ as the thermodynamic excess entropy of a system with potential energy $U(\bR)$ at the density $\rho$ of the configuration $\bR$:

\be\label{sex_def}
\Sex(\bR)\equiv \Sex(\rho,U(\bR))\,.
\ee
Here $\Sex(\rho,U)$ is the {\it thermodynamic} equilibrium (i.e., average) excess entropy of the state point with density $\rho$ and {\it average} potential energy $U$. Inverting this relation, the potential-energy function by definition obeys

\be\label{any}
U(\bR)=U(\rho,\Sex(\bR))\,
\ee
in which $U(\rho,\Sex)$ on the right-hand side is the thermodynamic equilibrium potential energy as a function of density $\rho$ and thermodynamic excess entropy $\Sex$, evaluated by substituting  $\Sex=\Sex(\bR)$. 

By definition, Eqs. (\ref{sex_def}) and (\ref{any}) apply for any system. We now limit the discussion to Roskilde-simple systems. Suppose $\bR_1$ is a configuration at density $\rho_1$ with the same reduced coordinate as $\bR_2$, a configuration at density $\rho_2$. It follows from Eq. (\ref{sex_def}) and the microcanonical expression for the excess entropy (Eq. (\ref{entr}) of the Appendix) that if ``Vol'' is the reduced-coordinate configuration-space volume, one has

\be\label{lign10}
\Sex(\bR_1)/k_B
\,=\, -N\ln N\,+\,\ln\left( {\rm Vol} \{\tbR'\,|\,U(\rho_1^{-1/3}\tbR')<U(\bR_1)\}\right)\,.
\ee
Likewise
\be
\Sex(\bR_2)/k_B
\,=\, -N\ln N\,+\,\ln\left( {\rm Vol} \{\tbR'\,|\,U(\rho_2^{-1/3}\tbR')<U(\bR_2)\}\right)\,.
\ee
Because $\bR_2=\rho_2^{-1/3}\rho_1^{1/3}\bR_1$, applying $\lambda=\rho_2^{-1/3}\rho_1^{1/3}$ in the $\Leftrightarrow$ version of Eq. (\ref{def}) to the inequality of the first set (Eq. (\ref{lign10})) we see that the two sets are identical. Thus $\Sex(\bR_1)=\Sex(\bR_2)$, which means that for a Roskilde-simple system $\Sex(\bR)$ depends only on the configuration's reduced coordinate:

\be\label{sextbr}
\Sex=\Sex(\tbR)\,.
\ee
Equation (\ref{any}) thus becomes

\be\label{fundeq}
U(\bR)=U(\rho,\Sex(\tbR))\,.
\ee
This ``$U=U$'' relation, which links the microscopic potential-energy function to the thermodynamic average potential-energy function, gives the fundamental characterization of Roskilde-simple systems. It is understood that, just as in the original isomorph theory, this identity is generally not obeyed exactly for all configurations, but to a good approximation for most of the physically relevant configurations. We proceed to derive the consequences of Eq. (\ref{fundeq}).

\subsection{Invariance of structure and dynamics along the configurational adiabats}\label{A}
 
In reduced coordinates Newton's second law for a system of identical masses is $d^2\tbR/d{\tilde t}^2=\tbF$ in which the reduced force vector is defined from the full force vector $\bF$ that give all particle forces in one single vector by $\tbF\equiv \bF \rho^{-1/3}/k_BT$ \cite{IV} (the below derivations all generalize straight away to systems of particles with different masses). 

In general one has $\tbF=\tbF(\bR)$, implying different dynamics at different state points. For a Roskilde-simple system, however, as we shall see now, the reduced force is a function of the reduced configuration vector, $\tbF=\tbF(\tbR)$. To show this, note that since $\nabla=\rho^{1/3}\tilde\nabla$, Eq. (\ref{fundeq}) implies $\bF=-\nabla U=-(\partial U/\partial \Sex)_\rho\,\rho^{1/3}\tilde\nabla \Sex(\tbR)$. Since $(\partial U/\partial \Sex)_\rho=T$, this means that $\tbF=-\tilde\nabla \Sex(\tbR)/k_B$. Thus the reduced force is a unique function of the reduced coordinates. This implies that the reduced-unit dynamics is invariant along the configurational adiabats, because via Eq. (\ref{sextbr}) two state points on a given isomorph -- given by a certain value of $\Sex$ -- correspond to the same range of reduced coordinate vectors $\tbR$.

The fact that the dynamics is invariant along the configurational adiabats immediately implies invariance of the reduced-unit structure and dynamics: If the same configurations are traced out in the course of time at two different state points -- except for a uniform scaling of space and time -- the structure as measured, e.g., via the reduced-unit radial distribution function or higher-order correlation functions must be identical at the two state points. This is of course consistent with the fact that $\Sex$, a measure of the structural disorder, by definition is constant along the configurational adiabats.

\subsection {Isomorphs}

Inspired by the above we {\it define} isomorphs as the configurational adiabats in the thermodynamic phase diagram. Thus {\it by definition} the excess entropy is an isomorph invariant and, as we have seen in Sec. \ref{A}, structure and dynamics are invariant to a good approximation along the isomorphs.

Expanding Eq. (\ref{fundeq}) to first order at constant density at any given state point leads to

\be\label{firstord}
U(\bR)\,\cong\, U\, +\, T(\rho,\Sex) \left(\Sex(\tbR)-\Sex\right)\,.
\ee
Consider two state points $(\rho_1,T_1)$ and $(\rho_2,T_2)$ with the same excess entropy $\Sex$. If $\bR_1$ and $\bR_2$ are two physically relevant configurations of these state points with same reduced coordinates, Eq. (\ref{firstord}) implies that if one for brevity writes $T(\rho_1,\Sex)=T_1$ and $T(\rho_2,\Sex)=T_2$, the following applies 

\be\label{isom}
\frac{U(\bR_1)-U_1}{k_B T_1}\,\cong\,\frac{U(\bR_2)-U_2}{k_B T_2}\,.
\ee
Changing sign and taking the exponential this becomes Eq. (\ref{isomeq}), the condition that the two state points are isomorphic according to the original definition \cite{IV}. Thus, as also stated in Sec. \ref{II}, the original formulation of the theory is the first-order approximation to the new formulation.

\subsection{Strong virial potential-energy correlations for constant-density fluctuations}

The microscopic virial is defined \cite{han13,tildesley} by 

\be\label{virdef}
W(\bR)\equiv -\frac{1}{3}\,\bR\cdot\nabla U(\bR)\,.
\ee
Recall that at any state point the average of $W(\bR)$ (denoted by $W$) gives the contribution to the pressure from the interactions via the general equation of state $pV=Nk_BT+W$ \cite{han13,tildesley}. 

Below, we first show that the potential energy determines the virial, which implies that these quantities are strongly correlated. Next we calculate the proportionality constant of the virial potential-energy fluctations. In regard to the first objective, suppose two configurations are given, $\bR_a$ and $\bR_b$, which have the same density and the same potential energy, $U(\bR_a)=U(\bR_b)$. We conclude from Eq. (\ref{altdef}) that $U(\lambda\bR_a)=U(\lambda\bR_b)$. Taking the derivative of this with respect to $\lambda$ results in $\bR_a\cdot\nabla U(\lambda\bR_a)=\bR_b\cdot\nabla U(\lambda\bR_b)$, which for $\lambda=1$ implies $W(\bR_a)=W(\bR_b)$. Thus any two configurations with same density and potential energy have the same virial. This means that $W$ is a function of $U$, which implies perfect correlations between potential energy and virial at constant density. No realistic systems obey Eq. (\ref{altdef}) perfectly, so in practice the correlations will not be perfect, but merely strong.

The constant of proportionality between the equilibrium virial and potential-energy  fluctuations at a given state point is denoted by $\gamma$ and referred to as the density-scaling exponent \cite{II,dyr14,IV}, i.e., $\gamma$ is characterized by 

\be\label{wucorr}
\Delta W(t)\cong \gamma\, \Delta U(t)\,.
\ee
Reference \onlinecite{IV} defined $\gamma$ at any given state point by

\be\label{gamma}
\gamma(\rho,\Sex)\equiv \left(\frac{\partial\ln T}{\partial\ln \rho}\right)_\Sex\,
\ee
and derived the general fluctuation expression 

\be\label{fluct}
\gamma(\rho,\Sex)\,=\,\frac{\langle\Delta W \Delta U\rangle}{\langle(\Delta U)^2\rangle}\,.
\ee
Here the angular brackets denote canonical $NVT$ averages. Whenever Eq. (\ref{wucorr}) is obeyed to a good approximation, i.e., for Roskilde-simple systems, Eq. (\ref{fluct}) implies that $\gamma$ of Eq. (\ref{gamma}) is the same as that appearing in Eq. (\ref{wucorr}), ensuring consistency. We proceed to derive Eq. (\ref{wucorr}) from Eq. (\ref{gamma}) for Roskilde-simple systems.

As shown in the Appendix $W(\bR)=\left({\partial U(\bR)}/{\partial\ln\rho}\right)_{\tbR}$ \cite{IV}, an expression which basically expresses that the virial is given by the work done to uniformly expand a given configuration. Substituting Eq. (\ref{fundeq}) into this expression leads to

\be\label{wrel}
W(\bR)=\left(\frac{\partial U(\rho,\Sex(\tbR))}{\partial\ln\rho}\right)_{\tbR}
=\left(\frac{\partial U(\rho,\Sex(\tbR))}{\partial\ln\rho}\right)_{\Sex}\,.
\ee
Expanding the partial derivative to first order around the equilibrium values of virial and excess entropy at the state point in question leads to

\be\label{wsexp}
W(\bR)\,\cong\,W\,+\left(\frac{\partial^2 U}{\partial\Sex\partial\ln\rho}\right) \left(\Sex(\tbR)-\Sex\right)\,.
\ee
Interchanging the orders of the differentiations and recalling that $\left(\partial U / \partial\Sex\right)_\rho=T$, we get

\be\label{WUrel0}
W(\bR)-W\cong\left(\frac{\partial T}{\partial\ln\rho}\right)_\Sex \left(\Sex(\tbR)-\Sex\right)\,.
\ee
Eliminating $\Sex(\tbR)-\Sex$ from Eqs. (\ref{WUrel0}) and (\ref{firstord}) leads to

\be\label{WUrel0a}
W(\bR)-W\cong\left(\frac{\partial\ln T}{\partial\ln\rho}\right)_\Sex \left(U(\tbR)-U\right)\,.
\ee
This can be rewritten as 

\be\label{WUrel}
W(\bR)\cong\gamma(\rho,\Sex)U(\bR)+C(\rho,\Sex)\,
\ee
in which $\gamma(\rho,\Sex)$ is the density-scaling exponent defined in Eq. (\ref{gamma}) and $C(\rho,\Sex)= W-\gamma(\rho,\Sex)U$. For the constant-density equilibrium fluctuations at a given state point Eq. (\ref{WUrel}) implies Eq. (\ref{wucorr}), i.e., strong virial potential-energy correlations.

\subsection{Single-parameter family of reduced-coordinate constant-potential-energy hypersurfaces}\label{D}

Molecular dynamics may be reformulated to deal with geodesic motion on the constant-potential-energy hypersurface, co-called {\it NVU} dynamics \cite{NVU_I, NVU_II}. In reduced coordinates, the constant-potential-energy hypersurfaces are the sets defined by $\{\tbR'\,|\,U(\rho^{-1/3}\tbR')={\rm Const.}\}$. In general, these sets are parameterized by the two parameters specifying a thermodynamic state point, e.g., density and temperature, corresponding to different dynamics at different state points. For a Roskilde-simple system, however, Eq. (\ref{fundeq}) implies that these sets are all constant-excess-entropy hypersurfaces. Consequently, the constant-potential-energy hypersurfaces are parameterized by a single number, the value of the excess entropy \cite{IV}. Each isomorph corresponds to a particular reduced-coordinate constant-potential-energy hypersurface, and the fact that these are parameterized by a single number throughout the two-dimensional thermodynamic phase diagram implies isomorph invariance of the {\it NVU} dynamics, which for most quantities give results that in the thermodynamic limit are identical to those of conventional Newtonian $NVT$ dynamics \cite{NVU_II}.

\section{Connecting to the approach of Sec. \ref{II}}\label{trehalv}

To establish the equivalence of the formulations of the new isomorph theory given in Secs. \ref{II} and \ref{III}, respectively, we consider two state points $(\rho_1,T_1)$ and $(\rho_2,T_2)$ with same excess entropy, $\Sex(\rho_1,T_1)=\Sex(\rho_2,T_2)$. If $\bR_1$ is a configuration corresponding to density $\rho_1$ and similarly for $\bR_2$, by definition of the microscopic entropy function (Eqs. (\ref{sex_def}) and (\ref{any})) we have
(recall that $U(\rho,\Sex)$ and $\Sex(\rho,U)$ are the thermodynamic functions relating state point averages)

\begin{eqnarray}
U(\bR_1)&=&U(\rho_1,\Sex(\rho_1,U(\bR_1)))\nonumber\\
U(\bR_2)&=&U(\rho_2,\Sex(\rho_2,U(\bR_2)))\,.
\end{eqnarray}
Writing as in Sec. \ref{II} for brevity $U(\bR_1)=U_1$ etc, if the two configurations have the same reduced coordinates, $\tbR_1=\tbR_2$, we have since Eq. (\ref{sextbr}) implies $\Sex(\rho_1,U_1)=\Sex(\rho_2,U_2)$

\be\label{U2lign}
U_2=U(\rho_2,\Sex(\rho_2,U_2))=U(\rho_2,\Sex(\rho_1,U_1))\,.
\ee
Comparing to Eq. (\ref{feq}) leads to the identification

\be\label{feqU}
f_1(\rho_2,U_1)=U(\rho_2,\Sex(\rho_1,U_1))\,.
\ee
To validate this expression we calculate the ratio $T_2/T_1$, which according to Sec. \ref{II} should be given by $T_2/T_1=(\partial f_1/\partial U_1)_{\rho_2}$. Since the two state points in question have same excess entropy, denoted by $\Sex$ below, and $\rho_1$ plays the role of reference density, i.e., is constant throughout, Eq. (\ref{feqU}) implies 

\be
\left(\frac{\partial f_1}{\partial U_1}\right)_{\rho_2}
=\left(\frac{\partial U(\rho_2,\Sex)}{\partial \Sex}\right)_{\rho_2}\left(\frac{\partial \Sex}{\partial U_1}\right)_{\rho_1}\,.
\ee
From the definition of temperature this gives the required

\be
\left(\frac{\partial f_1}{\partial U_1}\right)_{\rho_2}
=\,\frac{T_2}{T_1}\,.
\ee

\section{Concluding remarks}\label{fire}

Appendix A of the original isomorph paper Ref. \onlinecite{IV} showed that points B, C, and D of Sec. \ref{III} are equivalent, i.e., if any one of these three points applies, the two others follow by necessity. In that paper isomorphs were {\it defined} from the condition Eq. (\ref{isomeq}), and the reduced-unit isomorph invariance of structure and dynamics was {\it derived} from this equation. In Ref. \onlinecite{IV} Eq. (\ref{isomeq}) was {\it shown} to imply that isomorphs are configurational adiabats. In contrast, we have here {\it defined} the isomorphs as the configurational adiabats and {\it showed} that structure and dynamics are invariant along these. 

In the present treatment Eq. (\ref{isom}) and thus Eq. (\ref{isomeq}) is derived from a first-order expansion of the fundamental equation Eq. (\ref{fundeq}), which implies invariance of structure and dynamics when isomorphs are defined as the configurational adiabats. Moreover, the hidden-scale-invariance identity Eq. (\ref{hsi}) is replaced by Eq. (\ref{fundeq}), and the two abstract functions $h(\rho)$ and $\tPhi(\tbR)$ in Eq. (\ref{hsi}) are replaced by the temperature $T(\rho,S)$ and the microscopic entropy function $S(\tbR)$. In practice, the main changes compared to the original isomorph theory are that $C_V$ is only isomorph invariant to first order \cite{bai13} because the proof of its isomorph invariance was based on Eq. (\ref{isom}) \cite{IV} and that, likewise, the density-scaling exponent is only approximately a function merely of the density. Another implication is that the density-scaling phenomenon involves a hierarchy of approximations: In the simplest case the density scaling exponent $\gamma$ is constant, implying that the isomorphs (=isochrones) are given by $\rho^\gamma/T={\rm Const.}$ In the more general case described by the hidden scale invariance identity of the original formulation of the isomorph theory Eq. (\ref{hsi}), the isomorphs are given by  $h(\rho)/T={\rm Const.}$ corresponding to a density-scaling exponent (Eq. (\ref{gamma})) that may vary throughout the phase diagram, but only as a function of density. Finally, the present formulation allows for  the density-scaling exponent to vary more generally.

For a pair-potential system with $v(r)=\sum_n\varepsilon_n(r/\sigma)^{-n}$, because of the structural invariance along an isomorph in reduced coordinates one has $U(\rho,\Sex)=\sum_n C_n(\Sex)\rho^{n/3}$ \cite{boh12,ing12a}. This equation of state, which was previously derived by Rosenfeld assuming quasiuniversality \cite{ros82}, implies $T(\rho,\Sex)=\sum_n C'_n(\Sex)\rho^{n/3}$. For the LJ system this leads to the isomorph equation $\left[\alpha_{12}(\Sex)\rho^4-\alpha_6(\Sex)\rho^2\right]/T={\rm Const.}$, which implies the expressions tested numerically in Fig. 3. 

In summary, this paper proposes a new definition of a Roskilde-simple system, Eq. (\ref{def}). Equivalently, one may use Eq. (\ref{altdef}) as the definition. The original isomorph theory from 2009 \cite{IV} is recovered as the first-order approximation to the new one. The new definition does not change the class of Roskilde-simple systems. This class is still believed to include most van der Waals bonded and metallic solids and liquids, as well as the weakly ionic or dipolar systems, and exclude most hydrogen-bonded and covalently bonded systems, as well as the strongly ionic or dipolar systems. The new isomorph theory is simpler than the original one and its predictions are more accurate.

\appendix*
\section{Excess thermodynamics and the configuration-space microcanonical expression for the excess entropy}

We consider a system of $N$ identical particles in volume $V$ with density $\rho=N/V$. The particle coordinates are given by the $3N$-dimensional vector $\bR\equiv (\br_1,...,\br_N)$ and the corresponding reduced (dimensionless) coordinate vector is defined by $\tbR\equiv\rho^{1/3}\bR$. 
This Appendix derives an expression for the microscopic virial, summarizes the definition of excess (configurational) thermodynamic quantities, and derives the microcanonical expression for the excess entropy at the state point defined by density $\rho$ and average potential energy $U$:

\be\label{entr}
\Sex(\rho,U)/k_B
\,=\, -N\ln N\,+\,\ln\left( {\rm Vol} \{\tbR|U(\rho^{-1/3}\tbR)<U\}\right)\,.
\ee
Here ``Vol'' refers to the volume of the set in question, which is the $\tbR$ integral of the unity function over all configurations $\bR=\rho^{-1/3}\tbR$ with potential energy below $U$, i.e., $U(\bR)<U$.

\subsection{An expression for the microscopic virial}
Consider the infinitesimal uniform expansion $\bR\rightarrow (1+d\lambda)\bR$. The relative volume change is $dV/V=(1+d\lambda)^3-1=3\,d\lambda$, which implies $d\ln\rho=d\rho/\rho=-dV/V=-3\,d\lambda$, i.e., $d\lambda=-(1/3)\,d\ln\rho$. The change of the configuration vector is given by $d\bR=d\lambda\,\bR$, so the change of the potential energy is $dU(\bR)=d\lambda\,\bR\cdot\nabla U(\bR)=-(1/3)\,d\ln\rho\,\bR\cdot\nabla U(\bR)$. Comparing to the definition of the virial Eq. (\ref{virdef}) we get $dU(\bR)=d\ln\rho\, W(\bR)$. The reduced coordinate $\tbR$ is constant during the uniform expansion, so we conclude that

\be\label{W_Rtil}
W(\bR)
\,=\, \left(\frac{\partial U(\bR)}{\partial\ln\rho}\right)_{\tbR}\,.
\ee

\subsection{Excess thermodynamic quantities}

Recall from statistical mechanics that if the momentum degrees of freedom are denoted by $\bP\equiv (\bp_1,...,\bp_N)$ and $H(\bP,\bR)$ is the Hamiltonian, the Helmholtz free energy $F$ is given  by the classical partition function as follows (where $\beta\equiv 1/k_BT$) \cite{han13}

\be\label{fri}
e^{-\beta F}
\,=\, \frac{1}{N!}\int \frac{d\bP d\bR}{h^{3N}}\, e^{-\beta H(\bP,\bR)}\,.
\ee
The appearances of Planck's constant $h$ and the indistinguishability factor $1/N!$ ensure proper correspondence to quantum mechanics. These factors are conveniently absorbed by writing $F=\Fid+\Fex$ in which $\Fid$ is the free energy of an ideal gas at the same density and temperature, $\Fid=Nk_BT(\ln(\Lambda^3 \rho )-1)$ where $\Lambda=h/\sqrt{2\pi mk_BT}$ is the thermal de Broglie wavelength ($m$ is the particle mass) \cite{han13}. The result of these manipulations is that the excess free energy $\Fex$ is given by

\be\label{fri_conf}
e^{-\beta\Fex}
\,=\, \int \frac{d\bR}{V^{N}}\, e^{-\beta U(\bR)}\,.
\ee
In the case of free particles, $U=0$, we get $\Fex=0$ as required for consistency. Note that there is no $1/N!$ factor in Eq. (\ref{fri_conf}), so $\Fex$ is formally the free energy of a system of distinguishable particles with no momentum coordinates. 

Due to the separation $F=\Fid+\Fex$, all thermodynamic quantities that are derivatives of $F$ likewise separate into an ideal-gas contribution and an ``excess'' contribution. For instance, for the entropy one has $S=\Sid+\Sex$ in which $\Sex=-(\partial\Fex/\partial T)_\rho$, the isochoric specific heat separates into a sum of two terms and the well-known relation $\CVex=(\partial\Sex/\partial\ln T)_\rho$ applies, etc. 

The excess entropy obeys $\Sex<0$ because a liquid is more ordered than an ideal gas at the same density and temperature. As temperature goes to infinity, the system approaches the complete chaos of an ideal gas, so $\Sex\rightarrow 0$ for $T\rightarrow\infty$ at fixed density. The relation between excess entropy, potential energy, and temperature is the usual one, i.e.,

\be\label{temp}
\left(\frac{\partial\Sex}{\partial U}\right)_{\rho}
\,=\,\frac{1}{T}\,.
\ee
For the pressure the equation defining the average virial $W$, i.e., the average of $W(\bR)$ of Eq. (\ref{virdef}), $pV=Nk_BT+W$ \cite{han13}. This implies that $p=\pid+W/V$. Thus the excess pressure is $W/V$, which in terms of $\Fex$ is given by $W/V=-(\partial\Fex/\partial V)_T$.

\subsection{The microcanonical expression for the excess entropy}

The Heaviside theta function is denoted by $\Theta(x)$; recall that this function is unity for positive arguments and zero for negative,. The dimensionless volume of the set of configurations with potential energy less than $U$ is denoted by $\Omega(U)$ and given by 

\be\label{omega}
\Omega(U)
\,=\,\int \frac{d\bR}{V^{N}}\, \Theta(U-U(\bR))\,.
\ee
If $X_i$ is one of the $3N$ particle coordinates and $\partial_j\equiv\partial/\partial X_j$, the microcanonical average of $X_i\partial_j U(\bR)$ is by definition 

\be\label{mcav}
\langle X_i\partial_j U(\bR)\rangle_{\rm mc}
\,=\,  \frac{\int({d\bR}/{V^{N}})\,X_i(\partial_j U(\bR))\, \delta(U-U(\bR))}{\int({d\bR}/{V^{N}})\,\delta(U-U(\bR))} \,.
\ee
Following Pauli \cite{pauli}, via the fact that $\Theta'(x)=\delta(x)$ and a partial integration we get for the numerator

\begin{eqnarray}
\int\frac{d\bR}{V^{N}}\,X_i(\partial_j U(\bR))\, \delta(U-U(\bR))
&=&\frac{d}{dU}\int\frac{d\bR}{V^{N}}\,X_i(\partial_j U(\bR))\, \Theta(U-U(\bR))\nonumber\\
&=&\frac{d}{dU}\int\frac{d\bR}{V^{N}}\,X_i(\partial_j \left(U(\bR)-U\right))\,\Theta(U-U(\bR))\nonumber\\
&=&-\frac{d}{dU}\,\delta_{ij}\int\frac{d\bR}{V^{N}}\,\left(U(\bR)-U\right)\,\Theta(U-U(\bR))\nonumber\\
&=&\delta_{ij}\int\frac{d\bR}{V^{N}}\,\Theta(U-U(\bR))\nonumber\\
&=&\delta_{ij}\,\Omega(U)\,.
\end{eqnarray}
The denominator of Eq. (\ref{mcav}) is $\Omega'(U)$, so all together we get

\be\label{mcav2}
\langle X_i\partial_j U(\bR)\rangle_{\rm mc}
\,=\, \delta_{ij}\,\frac{\Omega(U)}{\Omega'(U)} \,.
\ee

Next, the canonical average of $X_i\partial_j U(\bR)$ is calculated. If $Z=\int{d\bR}/{V^{N}}\,\exp(-\beta U(\bR))$ is the partition function we have $\langle X_i\partial_j U(\bR)\rangle_{\rm can}=\int{d\bR}/{V^{N}}\,X_i(\partial_j U(\bR))\, \exp(-\beta U(\bR))/Z=-k_BT\int{d\bR}/{V^{N}}\,X_i\partial_j \exp(-\beta U(\bR))/Z$, which via a partial integration gives $k_BT\delta_{ij}$. Since averages are ensemble independent (in contrast to fluctuations), Eq. (\ref{mcav2}) implies 

\be\label{mcav3}
\frac{\Omega(U)}{\Omega'(U)}
\,=\,k_BT \,.
\ee
Combined with Eq. (\ref{temp}) this implies that $(\partial\Sex/\partial U)_\rho=1/T=k_B\, d\ln\Omega(U)/dU$, i.e.,

\be\label{Sex1}
\Sex
\,=\,k_B\ln\Omega(U)+{\rm Const.}
\ee
The constant is determined from the above-mentioned boundary condition $\Sex\rightarrow 0$ for $T\rightarrow\infty$ at constant density. From Eq. (\ref{omega}) we see that the constant is zero. Rewriting finally the definition of $\Omega(U)$ as an integral over the reduced coordinate vector $\tbR$ leads to Eq. (\ref{entr}).

\begin{acknowledgments}
The center for viscous liquid dynamics ``Glass and Time'' is sponsored by the Danish National Research Foundation via grant DNRF61.
\end{acknowledgments}


\begin{thebibliography}{42}
\expandafter\ifx\csname natexlab\endcsname\relax\def\natexlab#1{#1}\fi
\expandafter\ifx\csname bibnamefont\endcsname\relax
  \def\bibnamefont#1{#1}\fi
\expandafter\ifx\csname bibfnamefont\endcsname\relax
  \def\bibfnamefont#1{#1}\fi
\expandafter\ifx\csname citenamefont\endcsname\relax
  \def\citenamefont#1{#1}\fi
\expandafter\ifx\csname url\endcsname\relax
  \def\url#1{\texttt{#1}}\fi
\expandafter\ifx\csname urlprefix\endcsname\relax\def\urlprefix{URL }\fi
\providecommand{\bibinfo}[2]{#2}
\providecommand{\eprint}[2][]{\url{#2}}

\bibitem[{\citenamefont{Rice and Gray}(1965)}]{ric65}
\bibinfo{author}{\bibfnamefont{S.~A.} \bibnamefont{Rice}} \bibnamefont{and}
  \bibinfo{author}{\bibfnamefont{P.}~\bibnamefont{Gray}},
  \emph{\bibinfo{title}{{The Statistical Mechanics of Simple Liquids}}}
  (\bibinfo{publisher}{Interscience, New York}, \bibinfo{year}{1965}).

\bibitem[{\citenamefont{Temperley et~al.}(1968)\citenamefont{Temperley,
  Rowlinson, and Rushbrooke}}]{tem68}
\bibinfo{author}{\bibfnamefont{H.~N.~V.} \bibnamefont{Temperley}},
  \bibinfo{author}{\bibfnamefont{J.~S.} \bibnamefont{Rowlinson}},
  \bibnamefont{and} \bibinfo{author}{\bibfnamefont{G.~S.}
  \bibnamefont{Rushbrooke}}, \emph{\bibinfo{title}{{Physics of Simple
  Liquids}}} (\bibinfo{publisher}{Wiley, New York}, \bibinfo{year}{1968}).

\bibitem[{\citenamefont{Rowlinson and Widom}(1982)}]{rowlinson}
\bibinfo{author}{\bibfnamefont{J.~S.} \bibnamefont{Rowlinson}}
  \bibnamefont{and} \bibinfo{author}{\bibfnamefont{B.}~\bibnamefont{Widom}},
  \emph{\bibinfo{title}{{Molecular Theory of Capillarity}}}
  (\bibinfo{publisher}{Clarendon, Oxford}, \bibinfo{year}{1982}).

\bibitem[{\citenamefont{Chandler}(1987)}]{chandler}
\bibinfo{author}{\bibfnamefont{D.}~\bibnamefont{Chandler}},
  \emph{\bibinfo{title}{{Introduction to Modern Statistical Mechanics}}}
  (\bibinfo{publisher}{Oxford University Press}, \bibinfo{year}{1987}).

\bibitem[{\citenamefont{Barrat and Hansen}(2003)}]{barrat}
\bibinfo{author}{\bibfnamefont{J.-L.} \bibnamefont{Barrat}} \bibnamefont{and}
  \bibinfo{author}{\bibfnamefont{J.-P.} \bibnamefont{Hansen}},
  \emph{\bibinfo{title}{{Basic Concepts for Simple and Complex Liquids}}}
  (\bibinfo{publisher}{Cambridge University Press}, \bibinfo{year}{2003}).

\bibitem[{\citenamefont{Debenedetti}(2005)}]{deb05}
\bibinfo{author}{\bibfnamefont{P.~G.} \bibnamefont{Debenedetti}},
  \bibinfo{journal}{AICHE J.} \textbf{\bibinfo{volume}{51}},
  \bibinfo{pages}{2391} (\bibinfo{year}{2005}).

\bibitem[{\citenamefont{Bailey et~al.}(2008{\natexlab{a}})\citenamefont{Bailey,
  Pedersen, Gnan, Schr{\o}der, and Dyre}}]{I}
\bibinfo{author}{\bibfnamefont{N.~P.} \bibnamefont{Bailey}},
  \bibinfo{author}{\bibfnamefont{U.~R.} \bibnamefont{Pedersen}},
  \bibinfo{author}{\bibfnamefont{N.}~\bibnamefont{Gnan}},
  \bibinfo{author}{\bibfnamefont{T.~B.} \bibnamefont{Schr{\o}der}},
  \bibnamefont{and} \bibinfo{author}{\bibfnamefont{J.~C.} \bibnamefont{Dyre}},
  \bibinfo{journal}{J. Chem. Phys.} \textbf{\bibinfo{volume}{129}},
  \bibinfo{pages}{184507} (\bibinfo{year}{2008}{\natexlab{a}}).

\bibitem[{\citenamefont{Bailey et~al.}(2008{\natexlab{b}})\citenamefont{Bailey,
  Pedersen, Gnan, Schr{\o}der, and Dyre}}]{II}
\bibinfo{author}{\bibfnamefont{N.~P.} \bibnamefont{Bailey}},
  \bibinfo{author}{\bibfnamefont{U.~R.} \bibnamefont{Pedersen}},
  \bibinfo{author}{\bibfnamefont{N.}~\bibnamefont{Gnan}},
  \bibinfo{author}{\bibfnamefont{T.~B.} \bibnamefont{Schr{\o}der}},
  \bibnamefont{and} \bibinfo{author}{\bibfnamefont{J.~C.} \bibnamefont{Dyre}},
  \bibinfo{journal}{J. Chem. Phys.} \textbf{\bibinfo{volume}{129}},
  \bibinfo{pages}{184508} (\bibinfo{year}{2008}{\natexlab{b}}).

\bibitem[{\citenamefont{Bagchi and Chakravarty}(2010)}]{bag10}
\bibinfo{author}{\bibfnamefont{B.}~\bibnamefont{Bagchi}} \bibnamefont{and}
  \bibinfo{author}{\bibfnamefont{C.}~\bibnamefont{Chakravarty}},
  \bibinfo{journal}{J. Chem. Sci.} \textbf{\bibinfo{volume}{122}},
  \bibinfo{pages}{459} (\bibinfo{year}{2010}).

\bibitem[{\citenamefont{Hansen and McDonald}(2013)}]{han13}
\bibinfo{author}{\bibfnamefont{J.-P.} \bibnamefont{Hansen}} \bibnamefont{and}
  \bibinfo{author}{\bibfnamefont{I.~R.} \bibnamefont{McDonald}},
  \emph{\bibinfo{title}{{Theory of Simple Liquids: With Applications to Soft
  Matter}}} (\bibinfo{publisher}{Academic, New York}, \bibinfo{year}{2013}),
  \bibinfo{edition}{4th} ed.

\bibitem[{\citenamefont{Prasad and Chakravarty}(2014)}]{pra14}
\bibinfo{author}{\bibfnamefont{S.}~\bibnamefont{Prasad}} \bibnamefont{and}
  \bibinfo{author}{\bibfnamefont{C.}~\bibnamefont{Chakravarty}},
  \bibinfo{journal}{J. Chem. Phys.} \textbf{\bibinfo{volume}{140}},
  \bibinfo{eid}{164501} (\bibinfo{year}{2014}).

\bibitem[{\citenamefont{Dyre}(2014)}]{dyr14}
\bibinfo{author}{\bibfnamefont{J.~C.} \bibnamefont{Dyre}}, \bibinfo{journal}{J.
  Phys. Chem. B} \textbf{\bibinfo{volume}{118}}, \bibinfo{pages}{10007}
  (\bibinfo{year}{2014}).

\bibitem[{\citenamefont{Abramson}(2014)}]{abr14}
\bibinfo{author}{\bibfnamefont{E.~H.} \bibnamefont{Abramson}},
  \bibinfo{journal}{J. Phys. Chem. B} \textbf{\bibinfo{volume}{118}},
  \bibinfo{pages}{{(in press)}} (\bibinfo{year}{2014}).

\bibitem[{\citenamefont{Pond et~al.}(2011)\citenamefont{Pond, Errington, and
  Truskett}}]{pon11}
\bibinfo{author}{\bibfnamefont{M.~J.} \bibnamefont{Pond}},
  \bibinfo{author}{\bibfnamefont{J.~R.} \bibnamefont{Errington}},
  \bibnamefont{and} \bibinfo{author}{\bibfnamefont{T.~M.}
  \bibnamefont{Truskett}}, \bibinfo{journal}{J. Chem. Phys.}
  \textbf{\bibinfo{volume}{134}}, \bibinfo{pages}{081101}
  (\bibinfo{year}{2011}).

\bibitem[{\citenamefont{Malins et~al.}(2013)\citenamefont{Malins, Eggers, and
  Royall}}]{mal13}
\bibinfo{author}{\bibfnamefont{A.}~\bibnamefont{Malins}},
  \bibinfo{author}{\bibfnamefont{J.}~\bibnamefont{Eggers}}, \bibnamefont{and}
  \bibinfo{author}{\bibfnamefont{C.~P.} \bibnamefont{Royall}},
  \bibinfo{journal}{J. Chem. Phys.} \textbf{\bibinfo{volume}{139}},
  \bibinfo{pages}{234505} (\bibinfo{year}{2013}).

\bibitem[{\citenamefont{Flenner et~al.}(2014)\citenamefont{Flenner, Staley, and
  Szamel}}]{fle14}
\bibinfo{author}{\bibfnamefont{E.}~\bibnamefont{Flenner}},
  \bibinfo{author}{\bibfnamefont{H.}~\bibnamefont{Staley}}, \bibnamefont{and}
  \bibinfo{author}{\bibfnamefont{G.}~\bibnamefont{Szamel}},
  \bibinfo{journal}{Phys. Rev. Lett.} \textbf{\bibinfo{volume}{112}},
  \bibinfo{pages}{097801} (\bibinfo{year}{2014}).

\bibitem[{\citenamefont{Henao et~al.}(2014)\citenamefont{Henao, Pothoczki,
  Canales, Guardia, and Pardo}}]{hen14}
\bibinfo{author}{\bibfnamefont{A.}~\bibnamefont{Henao}},
  \bibinfo{author}{\bibfnamefont{S.}~\bibnamefont{Pothoczki}},
  \bibinfo{author}{\bibfnamefont{M.}~\bibnamefont{Canales}},
  \bibinfo{author}{\bibfnamefont{E.}~\bibnamefont{Guardia}}, \bibnamefont{and}
  \bibinfo{author}{\bibfnamefont{L.}~\bibnamefont{Pardo}}, \bibinfo{journal}{J.
  Mol. Liquids} \textbf{\bibinfo{volume}{190}}, \bibinfo{pages}{121}
  (\bibinfo{year}{2014}).

\bibitem[{\citenamefont{Pieprzyk et~al.}(2014)\citenamefont{Pieprzyk, Heyes,
  and Branka}}]{pie14}
\bibinfo{author}{\bibfnamefont{S.}~\bibnamefont{Pieprzyk}},
  \bibinfo{author}{\bibfnamefont{D.~M.} \bibnamefont{Heyes}}, \bibnamefont{and}
  \bibinfo{author}{\bibfnamefont{A.~C.} \bibnamefont{Branka}},
  \bibinfo{journal}{Phys. Rev. E} \textbf{\bibinfo{volume}{90}},
  \bibinfo{pages}{012106} (\bibinfo{year}{2014}).

\bibitem[{\citenamefont{Buchenau}(2014)}]{buc14}
\bibinfo{author}{\bibfnamefont{U.}~\bibnamefont{Buchenau}},
  \bibinfo{journal}{arXiv:1408.5767}  (\bibinfo{year}{2014}).

\bibitem[{\citenamefont{Fernandez and Lopez}(2014)}]{fer14}
\bibinfo{author}{\bibfnamefont{J.}~\bibnamefont{Fernandez}} \bibnamefont{and}
  \bibinfo{author}{\bibfnamefont{E.~R.} \bibnamefont{Lopez}},
  \emph{\bibinfo{title}{{{\rm in} Experimental Thermodynamics: Advances in
  Transport Properties of Fluids}}} (\bibinfo{publisher}{Royal Society of
  Chemistry}, \bibinfo{year}{2014}), Chap. \bibinfo{chapter}{9.3}, pp.
  \bibinfo{pages}{307--317}.

\bibitem[{\citenamefont{Schmelzer and Tropin}(2014)}]{sch14}
\bibinfo{author}{\bibfnamefont{J.~W.~P.} \bibnamefont{Schmelzer}}
  \bibnamefont{and} \bibinfo{author}{\bibfnamefont{T.~V.}
  \bibnamefont{Tropin}}, \bibinfo{journal}{J. Non-Cryst. Solids (in press)}
  (\bibinfo{year}{2014}).

\bibitem[{\citenamefont{Roland et~al.}(2005)\citenamefont{Roland,
  Hensel-Bielowka, Paluch, and Casalini}}]{rol05}
\bibinfo{author}{\bibfnamefont{C.~M.} \bibnamefont{Roland}},
  \bibinfo{author}{\bibfnamefont{S.}~\bibnamefont{Hensel-Bielowka}},
  \bibinfo{author}{\bibfnamefont{M.}~\bibnamefont{Paluch}}, \bibnamefont{and}
  \bibinfo{author}{\bibfnamefont{R.}~\bibnamefont{Casalini}},
  \bibinfo{journal}{Rep. Prog. Phys.} \textbf{\bibinfo{volume}{68}},
  \bibinfo{pages}{1405} (\bibinfo{year}{2005}).

\bibitem[{\citenamefont{Floudas et~al.}(2011)\citenamefont{Floudas, Paluch,
  Grzybowski, and Ngai}}]{flo11}
\bibinfo{author}{\bibfnamefont{G.}~\bibnamefont{Floudas}},
  \bibinfo{author}{\bibfnamefont{M.}~\bibnamefont{Paluch}},
  \bibinfo{author}{\bibfnamefont{A.}~\bibnamefont{Grzybowski}},
  \bibnamefont{and} \bibinfo{author}{\bibfnamefont{K.}~\bibnamefont{Ngai}},
  \emph{\bibinfo{title}{{Molecular Dynamics of Glass-Forming Systems: Effects
  of Pressure}}} (\bibinfo{publisher}{Springer, Berlin}, \bibinfo{year}{2011}).

\bibitem[{\citenamefont{Ngai et~al.}(2005)\citenamefont{Ngai, Casalini,
  Capaccioli, Paluch, and Roland}}]{nga05}
\bibinfo{author}{\bibfnamefont{K.~L.} \bibnamefont{Ngai}},
  \bibinfo{author}{\bibfnamefont{R.}~\bibnamefont{Casalini}},
  \bibinfo{author}{\bibfnamefont{S.}~\bibnamefont{Capaccioli}},
  \bibinfo{author}{\bibfnamefont{M.}~\bibnamefont{Paluch}}, \bibnamefont{and}
  \bibinfo{author}{\bibfnamefont{C.~M.} \bibnamefont{Roland}},
  \bibinfo{journal}{J. Phys. Chem. B} \textbf{\bibinfo{volume}{109}},
  \bibinfo{pages}{17356} (\bibinfo{year}{2005}).

\bibitem[{\citenamefont{Roed et~al.}(2013)\citenamefont{Roed, Gundermann, Dyre,
  and Niss}}]{roe13}
\bibinfo{author}{\bibfnamefont{L.~A.} \bibnamefont{Roed}},
  \bibinfo{author}{\bibfnamefont{D.}~\bibnamefont{Gundermann}},
  \bibinfo{author}{\bibfnamefont{J.~C.} \bibnamefont{Dyre}}, \bibnamefont{and}
  \bibinfo{author}{\bibfnamefont{K.}~\bibnamefont{Niss}}, \bibinfo{journal}{J.
  Chem. Phys.} \textbf{\bibinfo{volume}{139}}, \bibinfo{pages}{101101}
  (\bibinfo{year}{2013}).

\bibitem[{\citenamefont{Ubbelohde}(1965)}]{ubb65}
\bibinfo{author}{\bibfnamefont{A.~R.} \bibnamefont{Ubbelohde}},
  \emph{\bibinfo{title}{{Melting and Crystal Structure}}}
  (\bibinfo{publisher}{Clarendon, Oxford}, \bibinfo{year}{1965}).

\bibitem[{\citenamefont{Gnan et~al.}(2009)\citenamefont{Gnan, Schr{\o}der,
  Pedersen, Bailey, and Dyre}}]{IV}
\bibinfo{author}{\bibfnamefont{N.}~\bibnamefont{Gnan}},
  \bibinfo{author}{\bibfnamefont{T.~B.} \bibnamefont{Schr{\o}der}},
  \bibinfo{author}{\bibfnamefont{U.~R.} \bibnamefont{Pedersen}},
  \bibinfo{author}{\bibfnamefont{N.~P.} \bibnamefont{Bailey}},
  \bibnamefont{and} \bibinfo{author}{\bibfnamefont{J.~C.} \bibnamefont{Dyre}},
  \bibinfo{journal}{J. Chem. Phys.} \textbf{\bibinfo{volume}{131}},
  \bibinfo{pages}{234504} (\bibinfo{year}{2009}).

\bibitem[{\citenamefont{Dyre}(2013{\natexlab{a}})}]{dyr13}
\bibinfo{author}{\bibfnamefont{J.~C.} \bibnamefont{Dyre}},
  \bibinfo{journal}{Phys. Rev. E} \textbf{\bibinfo{volume}{87}},
  \bibinfo{pages}{022106} (\bibinfo{year}{2013}{\natexlab{a}}).

\bibitem[{\citenamefont{Pedersen}(2013)}]{ped13}
\bibinfo{author}{\bibfnamefont{U.~R.} \bibnamefont{Pedersen}},
  \bibinfo{journal}{J. Chem. Phys.} \textbf{\bibinfo{volume}{139}},
  \bibinfo{pages}{104102} (\bibinfo{year}{2013}).

\bibitem[{\citenamefont{Dyre}(2013{\natexlab{b}})}]{dyr13a}
\bibinfo{author}{\bibfnamefont{J.~C.} \bibnamefont{Dyre}},
  \bibinfo{journal}{Phys. Rev. E} \textbf{\bibinfo{volume}{88}},
  \bibinfo{pages}{042139} (\bibinfo{year}{2013}{\natexlab{b}}).

\bibitem[{\citenamefont{Bailey et~al.}(2013)\citenamefont{Bailey, B{\o}hling,
  Veldhorst, Schr\o{}der, and Dyre}}]{bai13}
\bibinfo{author}{\bibfnamefont{N.~P.} \bibnamefont{Bailey}},
  \bibinfo{author}{\bibfnamefont{L.}~\bibnamefont{B{\o}hling}},
  \bibinfo{author}{\bibfnamefont{A.~A.} \bibnamefont{Veldhorst}},
  \bibinfo{author}{\bibfnamefont{T.~B.} \bibnamefont{Schr\o{}der}},
  \bibnamefont{and} \bibinfo{author}{\bibfnamefont{J.~C.} \bibnamefont{Dyre}},
  \bibinfo{journal}{J. Chem. Phys.} \textbf{\bibinfo{volume}{139}},
  \bibinfo{pages}{184506} (\bibinfo{year}{2013}).

\bibitem[{rum()}]{rumd}
\emph{\bibinfo{title}{All simulations were performed using a molecular dynamics
  code optimized for \textit{NVIDIA} graphics cards, which is available as open
  source code at http://rumd.org.}}

\bibitem[{\citenamefont{B{\o}hling et~al.}(2012)\citenamefont{B{\o}hling,
  Ingebrigtsen, Grzybowski, Paluch, Dyre, and Schr{\o}der}}]{boh12}
\bibinfo{author}{\bibfnamefont{L.}~\bibnamefont{B{\o}hling}},
  \bibinfo{author}{\bibfnamefont{T.~S.} \bibnamefont{Ingebrigtsen}},
  \bibinfo{author}{\bibfnamefont{A.}~\bibnamefont{Grzybowski}},
  \bibinfo{author}{\bibfnamefont{M.}~\bibnamefont{Paluch}},
  \bibinfo{author}{\bibfnamefont{J.~C.} \bibnamefont{Dyre}}, \bibnamefont{and}
  \bibinfo{author}{\bibfnamefont{T.~B.} \bibnamefont{Schr{\o}der}},
  \bibinfo{journal}{New J. Phys.} \textbf{\bibinfo{volume}{14}},
  \bibinfo{pages}{113035} (\bibinfo{year}{2012}).

\bibitem[{\citenamefont{Ingebrigtsen et~al.}(2012)\citenamefont{Ingebrigtsen,
  B{\o}hling, Schr{\o}der, and Dyre}}]{ing12a}
\bibinfo{author}{\bibfnamefont{T.~S.} \bibnamefont{Ingebrigtsen}},
  \bibinfo{author}{\bibfnamefont{L.}~\bibnamefont{B{\o}hling}},
  \bibinfo{author}{\bibfnamefont{T.~B.} \bibnamefont{Schr{\o}der}},
  \bibnamefont{and} \bibinfo{author}{\bibfnamefont{J.~C.} \bibnamefont{Dyre}},
  \bibinfo{journal}{J. Chem. Phys.} \textbf{\bibinfo{volume}{136}},
  \bibinfo{pages}{061102} (\bibinfo{year}{2012}).

\bibitem[{\citenamefont{Barros~de Oliveira et~al.}(2006)\citenamefont{Barros~de
  Oliveira, Netz, Colla, and Barbosa}}]{deo06}
\bibinfo{author}{\bibfnamefont{A.}~\bibnamefont{Barros~de Oliveira}},
  \bibinfo{author}{\bibfnamefont{P.~A.} \bibnamefont{Netz}},
  \bibinfo{author}{\bibfnamefont{T.}~\bibnamefont{Colla}}, \bibnamefont{and}
  \bibinfo{author}{\bibfnamefont{M.~C.} \bibnamefont{Barbosa}},
  \bibinfo{journal}{J. Chem. Phys.} \textbf{\bibinfo{volume}{124}},
  \bibinfo{eid}{084505} (\bibinfo{year}{2006}).

\bibitem[{\citenamefont{Landau and Lifshitz}(1958)}]{lan58}
\bibinfo{author}{\bibfnamefont{L.~D.} \bibnamefont{Landau}} \bibnamefont{and}
  \bibinfo{author}{\bibfnamefont{E.~M.} \bibnamefont{Lifshitz}},
  \emph{\bibinfo{title}{Statistical Physics}} (\bibinfo{publisher}{Pergamon,
  Oxford}, \bibinfo{year}{1958}).

\bibitem[{\citenamefont{Rovelli}(2014)}]{rov14}
\bibinfo{author}{\bibfnamefont{C.}~\bibnamefont{Rovelli}},
  \bibinfo{journal}{arXiv:1407.3384}  (\bibinfo{year}{2014}).

\bibitem[{\citenamefont{Allen and Tildesley}(1987)}]{tildesley}
\bibinfo{author}{\bibfnamefont{M.~P.} \bibnamefont{Allen}} \bibnamefont{and}
  \bibinfo{author}{\bibfnamefont{D.~J.} \bibnamefont{Tildesley}},
  \emph{\bibinfo{title}{{Computer Simulation of Liquids}}}
  (\bibinfo{publisher}{Oxford Science Publications}, \bibinfo{year}{1987}).

\bibitem[{\citenamefont{Ingebrigtsen
  et~al.}(2011{\natexlab{a}})\citenamefont{Ingebrigtsen, Toxvaerd, Heilmann,
  Schr{\o}der, and Dyre}}]{NVU_I}
\bibinfo{author}{\bibfnamefont{T.~S.} \bibnamefont{Ingebrigtsen}},
  \bibinfo{author}{\bibfnamefont{S.}~\bibnamefont{Toxvaerd}},
  \bibinfo{author}{\bibfnamefont{O.~J.} \bibnamefont{Heilmann}},
  \bibinfo{author}{\bibfnamefont{T.~B.} \bibnamefont{Schr{\o}der}},
  \bibnamefont{and} \bibinfo{author}{\bibfnamefont{J.~C.} \bibnamefont{Dyre}},
  \bibinfo{journal}{J. Chem. Phys.} \textbf{\bibinfo{volume}{135}},
  \bibinfo{pages}{104101} (\bibinfo{year}{2011}{\natexlab{a}}).

\bibitem[{\citenamefont{Ingebrigtsen
  et~al.}(2011{\natexlab{b}})\citenamefont{Ingebrigtsen, Toxvaerd, Schr{\o}der,
  and Dyre}}]{NVU_II}
\bibinfo{author}{\bibfnamefont{T.~S.} \bibnamefont{Ingebrigtsen}},
  \bibinfo{author}{\bibfnamefont{S.}~\bibnamefont{Toxvaerd}},
  \bibinfo{author}{\bibfnamefont{T.~B.} \bibnamefont{Schr{\o}der}},
  \bibnamefont{and} \bibinfo{author}{\bibfnamefont{J.~C.} \bibnamefont{Dyre}},
  \bibinfo{journal}{J. Chem. Phys.} \textbf{\bibinfo{volume}{135}},
  \bibinfo{pages}{104102} (\bibinfo{year}{2011}{\natexlab{b}}).

\bibitem[{\citenamefont{Rosenfeld}(1982)}]{ros82}
\bibinfo{author}{\bibfnamefont{Y.}~\bibnamefont{Rosenfeld}},
  \bibinfo{journal}{Phys. Rev. A} \textbf{\bibinfo{volume}{26}},
  \bibinfo{pages}{3633} (\bibinfo{year}{1982}).

\bibitem[{\citenamefont{Pauli}(1973)}]{pauli}
\bibinfo{author}{\bibfnamefont{W.}~\bibnamefont{Pauli}},
  \emph{\bibinfo{title}{{Pauli Lectures on Physics. Volume 4. Statistical
  Mechanics}}} (\bibinfo{publisher}{MIT Press, Cambridge, Massachusetts},
  \bibinfo{year}{1973}).

\end{thebibliography}
\end{document}